\documentclass[sigconf]{acmart}

\AtBeginDocument{%
  }

\copyrightyear{2025}
\acmYear{2025}
\setcopyright{acmlicensed}\acmConference[KDD '25]{Proceedings of the 31st ACM SIGKDD Conference on Knowledge Discovery and Data Mining V.1}{August 3--7, 2025}{Toronto, ON, Canada}
\acmBooktitle{Proceedings of the 31st ACM SIGKDD Conference on Knowledge Discovery and Data Mining V.1 (KDD '25), August 3--7, 2025, Toronto, ON, Canada}
\acmDOI{10.1145/3690624.3709253}
\acmISBN{979-8-4007-1245-6/25/08}

\usepackage[utf8]{inputenc} 
\usepackage[T1]{fontenc}    
\usepackage{hyperref}       
\usepackage{url}            
\usepackage{booktabs}       
\usepackage{amsfonts}       
\usepackage{nicefrac}       
\usepackage{microtype}      
\usepackage{xcolor}         

\usepackage{float}
\usepackage{graphicx}
\usepackage{subfigure}
\usepackage{wrapfig} 
\usepackage{booktabs}
\usepackage{colortbl} 
\usepackage{amsmath}

\newcommand{\rmh}{\mathbf{h}}

\newcommand{\rmx}{\mathbf{x}}

\newcommand{\rmp}{\mathbf{p}}

\newcommand{\method}{FABind+}

\begin{document}

\title{\method{}: Enhancing Molecular Docking through \\ Improved Pocket Prediction and Pose Generation}

\author{Kaiyuan Gao}
\email{im_kai@hust.edu.cn}
\orcid{0009-0002-8862-8320}
\affiliation{%
  \institution{Huazhong University of Science and Technology}
  \city{Wuhan}
  \state{Hubei}
  \country{China}
}

\author{Qizhi Pei}
\email{qizhipei@ruc.edu.cn}
\orcid{0000-0002-7242-422X}
\affiliation{%
  \institution{Renmin University of China}
  \city{Beijing}
  \country{China}
}

\author{Gongbo Zhang}
\email{gary_z@hust.edu.cn}
\orcid{0009-0007-7193-7967}
\affiliation{%
  \institution{Huazhong University of Science and Technology}
  \city{Wuhan}
  \state{Hubei}
  \country{China}
}

\author{Jinhua Zhu}
\email{teslazhu@mail.ustc.edu.cn}
\orcid{0000-0003-2157-9077}
\affiliation{%
  \institution{University of Science and Technology of China}
  \city{Hefei}
  \state{Anhui}
  \country{China}
}

\author{Kun He*}
\email{brooklet60@hust.edu.cn}
\orcid{0000-0001-7627-4604}
\affiliation{%
  \institution{Huazhong University of Science and Technology}
  \city{Wuhan}
  \state{Hubei}
  \country{China}
}

\author{Lijun Wu}
\authornote{Corresponding authors.}
\email{lijun_wu@outlook.com}
\orcid{0000-0002-3530-590X}
\affiliation{%
  \institution{Shanghai Artificial Intelligence Laboratory}
  \city{Shanghai}
  \country{China}  
}

\renewcommand{\shortauthors}{Kaiyuan Gao et al.}

\begin{abstract}
Molecular docking is a pivotal process in drug discovery. While traditional techniques rely on extensive sampling and simulation governed by physical principles, deep learning has emerged as a promising alternative, offering improvements in both accuracy and efficiency. 
Building upon the foundational work of FABind, a model focused on speed and accuracy, we introduce \method{}, an enhanced iteration that significantly elevates the performance of its predecessor. We identify pocket prediction as a critical bottleneck in molecular docking and introduce an enhanced approach. In addition to the pocket prediction module, the docking module has also been upgraded with permutation loss and a more refined model design. These designs enable the regression-based \method{} to surpass most of the generative models.
In contrast, while sampling-based models often struggle with inefficiency, they excel in capturing a wide range of potential docking poses, leading to better overall performance. To bridge the gap between sampling and regression docking models, we incorporate a simple yet effective sampling technique coupled with a lightweight confidence model, transforming the regression-based \method{} into a sampling version without requiring additional training. This involves the introduction of pocket clustering to capture multiple binding sites and dropout sampling for various conformations. The combination of a classification loss and a ranking loss enables the lightweight confidence model to select the most accurate prediction.
Experimental results and analysis demonstrate that \method{} (both the regression and sampling versions) not only significantly outperforms the original FABind, but also achieves competitive state-of-the-art performance. Our code is available at \url{https://github.com/QizhiPei/FABind}. 

\end{abstract}




\maketitle

\section{Introduction}


\begin{figure}
    \centering
    \includegraphics[width=0.85\linewidth]{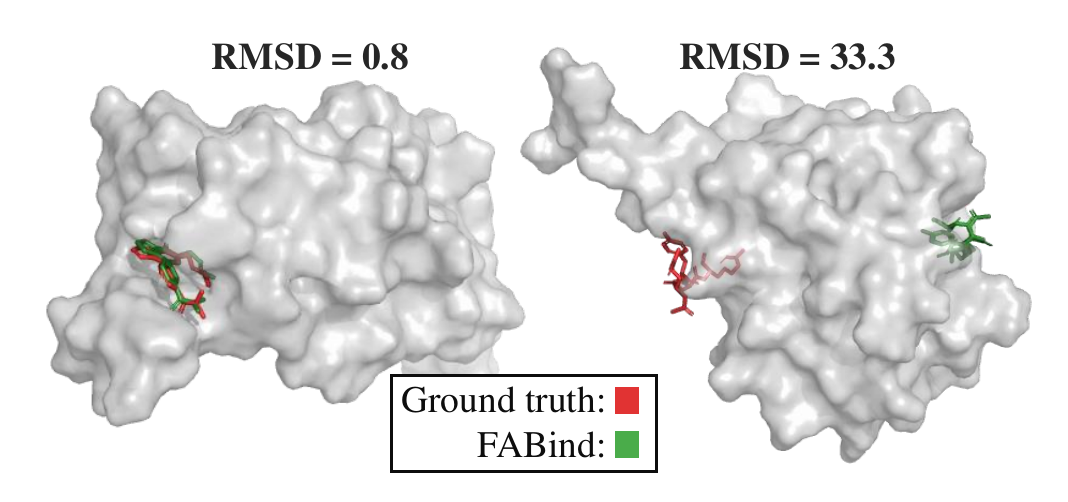}
    \caption{Pocket prediction is critical for docking. \textit{Left:} a good case with correct pocket and docking pose prediction. \textit{Right:} a bad case with incorrect pocket prediction.}
    \label{fig:pocket_showcase_intro}
    \vspace{-0.3cm}
\end{figure}

Molecular docking is a foundational technique in drug discovery~\cite{morris1996distributed,morris2008molecular,thomsen2006moldock,ai4science2023impact}, predicting the preferred orientation of ligands when bound to a protein target. This process is crucial for the identification and optimization of compounds with therapeutic potential. Traditional methodologies~\cite{friesner2004glide,trott2010autodock,morris1996distributed,verdonk2003improved} rely heavily on exhaustive sampling and simulation techniques based on the principles of physics and chemistry~\cite{shoichet1993matching,bernacki2005virtual,perez2016advances,huang2007physics}. These approaches aim to mimic the complex interactions between ligands and protein receptors to forecast optimal conformations. However, despite their extensive application, the classical methods are often criticized for their computational intensity, causing slow processing and significant resource consumption~\cite{pagadala2017software,huang2006molecular}. 

With the rise of computational power and machine learning, deep learning approaches for molecular docking have emerged recently~\cite{zhang2023efficient,lu2023dynamicbind,zhou2023uni}. These deep learning methods can be broadly categorized into two types: regression approaches and generative models. Regression approaches directly predict the coordinates~\cite{stark2022equibind,zhang2022e3bind} or distance matrices of pocket-ligand interactions~\cite{masters2022deep,lu2022tankbind}. In contrast, generative models sample multiple candidate poses and use confidence models to select the most promising ones~\cite{corso2023diffdock,wang2023flexidock,guo2023diffdock}, similar to traditional methods. 
FABind, a notable example of a regression-based method, has demonstrated significant advancements in terms of speed and accuracy. However, despite its impressive performance, FABind's accuracy does not fully align with the latest advancements in the field~\cite{corso2023diffdock,liu2023pre,yan2023multi}.

\begin{figure*}
\centering
\includegraphics[width=0.86\linewidth]{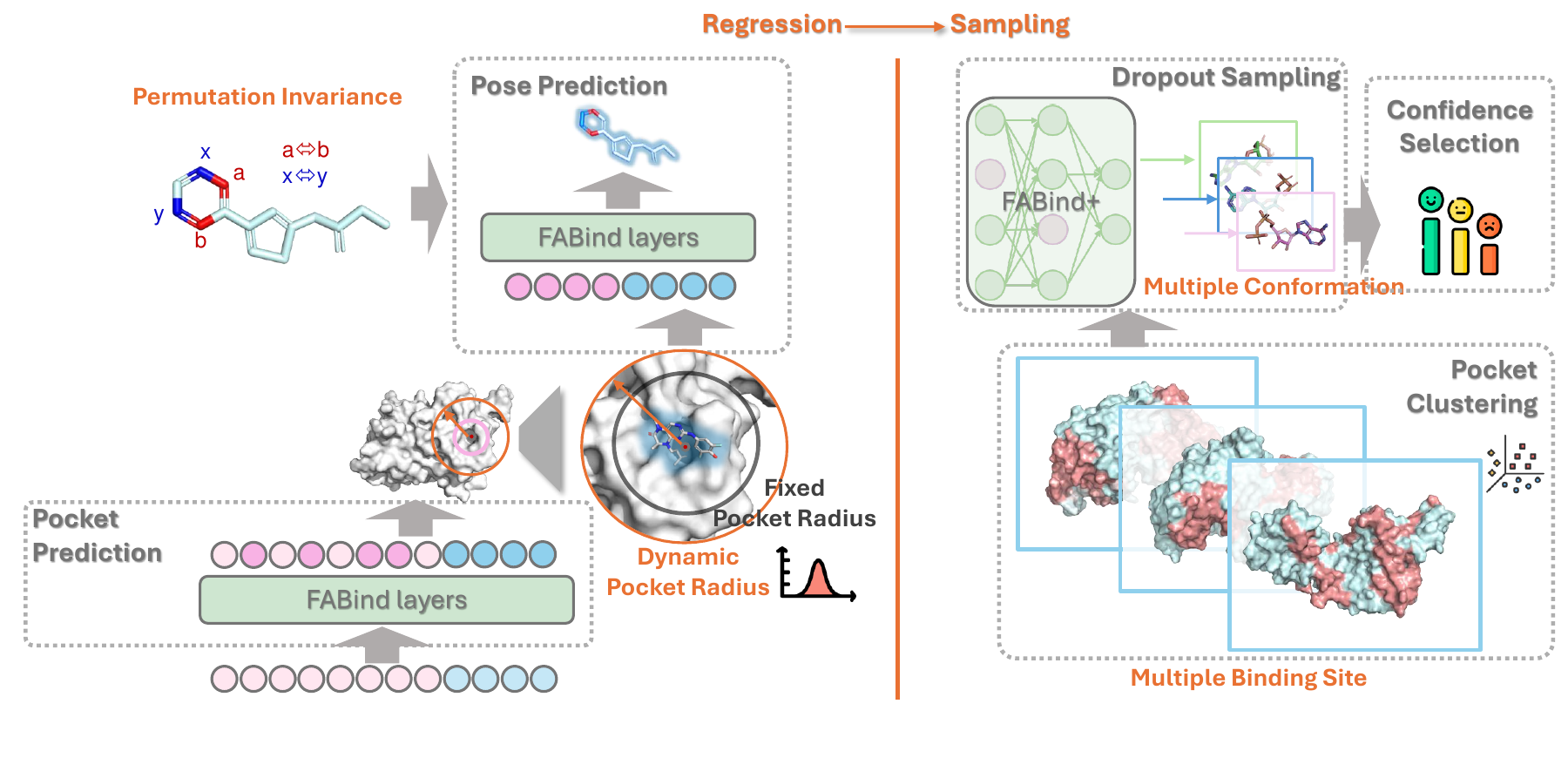}
\vspace{-0.5cm}
\caption{Overall framework of \method{}. \textit{Left:} Pipeline of FABind with the newly proposed dynamic pocket radius prediction and permutation loss module. The protein-ligand complex graph, with pink and blue nodes representing protein and ligand atoms respectively, is first processed by the pocket prediction module to identify the binding pocket. The identified pocket (highlighted in dark pink) is then utilized for docking in the pose prediction module. \textit{Right:} The sampling version of \method{}, which contains a pocket clustering module, a conformation sampling module, and a confidence selection module.}
\label{fig:overview}
\end{figure*}

In this work, we introduce \method{}, which features enhanced pocket prediction capabilities and high-quality pose generation.
Our initial analysis identifies pocket prediction as a key factor influencing docking accuracy, where inaccuracies can lead to suboptimal results. As illustrated in Figure~\ref{fig:pocket_showcase_intro}, incorrect pocket predictions from FABind result in a significant root-mean-square deviation (RMSD), with 16.25\% of poses exhibiting an RMSD greater than 10{\AA}, largely due to these prediction errors.
To mitigate this, we propose dynamically adjusting the pocket radius instead of relying on a fixed-size sphere, which allows the pocket to better encompass the entire potential ligand structure, thereby improving docking pose predictions. 
For the docking module, we also implement strategies to improve the performance.  Inspired by previous works on molecular structure modeling~\cite{dmcg,zhang2022ligpose}, which emphasize the importance of permutation invariance for symmetric atoms, we introduce a permutation loss function to increase the robustness of conformation predictions. Furthermore, we conduct model adjustments to optimize training and enhance overall performance.

Demonstrated by DiffDock and subsequent works~\cite{corso2023discovery,corso2023diffdock}, a sampling model is essential for capturing multiple binding sites and conformations, leading to better docking results. However, previous sampling-based methods that train generative models, mostly diffusion-based, incur high inference costs, which are undesirable in high-throughput virtual screening scenarios. Unlike previous approaches, we innovatively transform the pre-trained regression-based \method{} into a sampling-based model without further training while maintaining the same inference time for one sampling. 
Specifically, to enable discovering multiple binding sites, we employ a clustering method to identify all potential pocket candidates, leveraging our residue-level probability for pockets. 
Besides, we integrate a simple yet effective sampling mechanism based on dropout~\cite{srivastava2014dropout} to enable \method{} sample multiple conformations. Finally, a lightweight confidence model is trained on the fly to select the best-sampled structure. We additionally emphasize the importance of loss design and demonstrate that a lightweight model architecture is sufficient for docking pose selection.


For evaluation, we conduct comprehensive experiments and analyses on the widely recognized PDBBind v2020 benchmarks~\cite{liu2017forging} under various settings. As a regression-based model, \method{} outperforms all previous methods with remarkable inference speed compared to its generative counterparts. Additionally, by activating the sampling mode, \method{} is capable of generating diverse and high-quality conformations, capturing multiple pockets and multiple conformations that the regression model alone might miss, while also achieving improved docking accuracy through confidence model selection.
\section{Related Work}
\noindent\textbf{Pocket/Binding Site Prediction.}
Pocket prediction plays a key role in structure-based drug discovery. Early computational methods rely on hand-crafted features and use different modeling approaches~\cite{stank2016protein,weisel2007pocketpicker,capra2009predicting}. 
For example, sequence-based techniques exploited protein sequences~\cite{taherzadeh2016sequence,chen2009sequence}, while structure-based tools examined 3D structures~\cite{le2009fpocket,laskowski1995surfnet}, and the integration of both sequence and 3D structures~\cite{capra2009predicting}.
Recently, deep learning has achieved significant advances in this area. Most contemporary methods, utilizing voxel-based~\cite{viart2020pickpocket,pu2019deepdrug3d,jimenez2017deepsite,ragoza2017protein} and node-based representations~\cite{le2009fpocket,krivak2018p2rank,zhang2023equipocket}, implement dynamic pocket prediction, where larger pockets are predicted for larger ligands. Contrastively, Fixed pocket radius in TankBind~\cite{lu2022tankbind}, E3Bind~\cite{zhang2022e3bind} and FABind~\cite{pei2023fabind} is not sufficient. Moreover, it is crucial to model multiple binding sites within a single protein, as previous work has demonstrated~\cite{le2009fpocket,krivak2018p2rank}.


\noindent\textbf{Molecular Docking.}
Molecular docking predicts the correct binding pose of protein-ligand complexes. Traditional methods are typically sampling-based, which involves optimizing various initial conformations to generate different binding structures~\cite{trott2010autodock,koes2013lessons,friesner2004glide,jones1997development}. Geometric deep learning has shown promise in docking prediction, divided into two categories: (1) Regression-based methods directly predict the docked ligand pose coordinates or optimize the structures with predicted pairwise distances between atoms, such as EquiBind~\cite{stark2022equibind}, TankBind~\cite{lu2022tankbind}, E3Bind~\cite{zhang2022e3bind}, and KarmaDock~\cite{zhang2023efficient}. These methods usually demonstrate clear advantages in inference speed. (2) Sampling-based methods require multiple ligand poses sampling and then perform optimization or selection among sampled conformation candidates~\cite{nakata2023end,corso2023diffdock,wang2023flexidock,guo2023diffdock}. While more computationally expensive, sampling-based methods often yield more accurate predictions. Notably, Alphafold 3~\cite{abramson2024af3} achieves huge breakthroughs, but is not open-sourced and lacks detailed methodological transparency. DeltaDock~\cite{yan2023multi} and HelixDock~\cite{liu2023pre} also obtain competitive results. However, they either generate large-scale data with simulators or add extensive high-quality data for the training, which is not a fair comparison. Among related works, FABind is a regression-based approach with both efficiency and effectiveness. Our work follows FABind to enhance regression results and unlock its sampling potential.

\section{Methodology}
\subsection{Preliminaries}
\noindent\textbf{Problem Definition.} 
Let $\mathcal{G}=(\mathcal{V}:=\{\mathcal{V}^l, \mathcal{V}^p\},\mathcal{E}:=\{\mathcal{E}^l, \mathcal{E}^p, \mathcal{E}^{lp}\})$ denotes a protein-ligand complex, where $\mathcal{G}^l=(\mathcal{V}^l, \mathcal{E}^l)$ and $\mathcal{G}^p=(\mathcal{V}^p, \mathcal{E}^p)$ are ligand and protein graph, respectively. The symbols $\mathcal{V}$ and $\mathcal{E}$ represent collections of atoms (residue for protein) and bonds. $\mathcal{E}^{lp}$ in $\mathcal{E}$ is the edge collection between protein and ligand graph. Each node $v=(\mathbf{h}, \mathbf{x})\in\mathcal{V}$ contains a feature vector $\mathbf{h}$ and its coordinates $\mathbf{x}\in\mathbb{R}^{3}$. For clarity, $v_i=(\mathbf{h}_i, \mathbf{x}_i)\in\mathcal{V}^l$ is used to denote ligand atom and $v_j=(\mathbf{h}_j, \mathbf{x}_j)\in\mathcal{V}^p$ for protein residue. We use $\mathcal{R}^l \in \mathbb{R}^{\mid \mathcal{V}^l \mid \times 3}$ and $\mathcal{R}^p \in \mathbb{R}^{\mid \mathcal{V}^p \mid \times 3}$ to represent the conformation of ligand and protein, respectively. Given a protein in its bound state and a flexible ligand, our objective is to learn a mapping from the randomly initialized ligand pose to the bounded conformation $\mathcal{R}^l=\{\mathbf{x}_i\}_{1\leq i \leq \mid \mathcal{V}^l \mid}$.
Notably, we focus on the blind docking setting, where we possess no information regarding the binding pocket. 

\noindent\textbf{Key Idea of FABind.}
FABind~\cite{pei2023fabind} is a novel deep learning framework that aims to provide both fast and accurate protein-ligand binding structure prediction in an end-to-end manner. It seeks to overcome the inefficiencies typically associated with sampling methods and the inaccuracies often seen in regression-based approaches. The first core innovation in FABind is its unique layer design called ``\textbf{F}ABind layer" (F). Each FABind layer $(l)$ processes the protein-ligand complex graph and the protein-ligand pair embedding, producing updated embeddings and ligand structures:
\begin{equation}\small
\label{eq:fabind_layer}
\rmh_i^{(l+1)}, \rmh_j^{(l+1)}, \rmx_i^{(l+1)}, \rmp_{ij}^{(l+1)} =\nonumber\text{F}(\rmh_i^{(l)}, \rmh_j^{(l)}, \rmx_i^{(l)}, \rmx_j, \rmp_{ij}^{(l)}), \nonumber
\end{equation}
where $\rmp_{ij} \in \mathbb{R}^{D\times D}$ is the pair embedding for each pair of ligand and protein node and $D$ is the hidden size.
The FABind layer contains three key components: independent message passing, cross-attention update, and interfacial message passing. The independent message passing captures interactions within the protein and ligand separately, the cross-attention update enhances node representations by exchanging information across the protein and ligand, and the interfacial message passing focuses on modeling interactions at the protein-ligand interface. 

Another important contribution of FABind is the decomposition of the blind docking process into pocket prediction and pocket-specific docking. The unified framework introduces a ligand-informed pocket prediction module that utilizes ligand information to identify the unique binding pocket. This approach achieves faster and more precise pocket prediction compared to methods that rely on external pocket detectors. The predicted pocket is then used for downstream docking prediction.

\subsection{Enhancing FABind to \method{}}
Though FABind achieves comparable docking performance in a highly efficient manner, it does not demonstrate significant advantages in docking accuracy over existing works~\cite{corso2023diffdock,yan2023multi}. Our detailed evaluation of the predicted docking poses reveals that the limitations of FABind stem from both inaccurate pocket predictions and the capabilities of the docking module. Therefore, we propose 
new approaches to improve both the pocket prediction and the docking modules to enhance final docking.
The overall framework of regression \method{} is shown as the left part in Figure~\ref{fig:overview}.

\subsubsection{Pocket Prediction}
\begin{figure}
    \centering
    \includegraphics[width=0.7\linewidth]{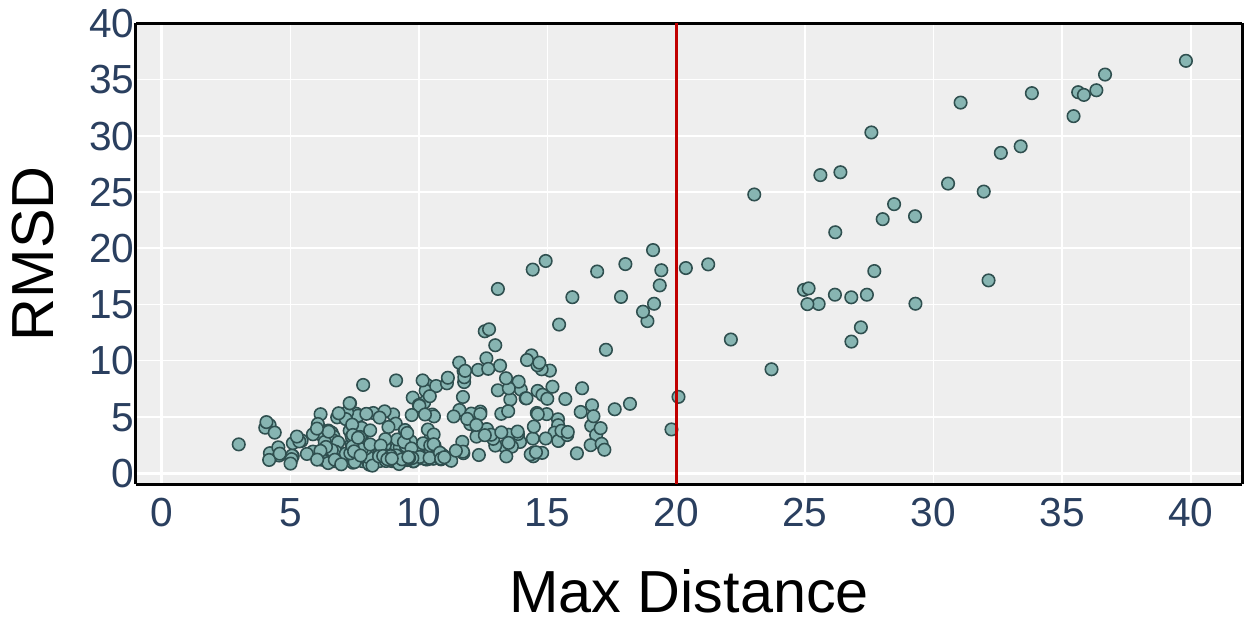}
    \vspace{-0.3cm}
    \caption{Analysis of pocket prediction and predicted ligand RMSD. ``Max Distance'' is the distance between the predicted pocket center and the farthest ground truth ligand atom.}
    \label{fig:max_dist}
    \vspace{-0.3cm}
\end{figure}

Regarding pocket prediction, we found that FABind was easy to predict incorrect pocket positions, resulting in bad docking pose prediction for the next stage, as illustrated in Figure~\ref{fig:pocket_showcase_intro}. Therefore, we attempt to improve the pocket prediction method in this work. 
While FABind effectively predicts the pocket center, it uses a fixed-radius sphere to define the overall pocket, leading to inaccuracies. If the fixed radius is too small, the pocket fails to encompass all possible amino acids and ligand atoms. Since our docking network requires interactions between all ligand atoms and pocket amino acids, ensuring the pocket's size can contain all ligand atoms is crucial for the docking module's performance.

To verify the above hypothesis, we plot the max distance between the predicted pocket center (with coordinates $\mathbf{x}^p_o$) and each ground truth ligand atom (with coordinates $\mathbf{x}^l_i$, where $1\leq i \leq|\mathcal{V}^l|$). The maximum distance, denoted as $D_{max}$, is formulated as: $D_{\text{max}} = \max_{1 \leq i \leq |\mathcal{V}^l|} \|\mathbf{x}^p_o - \mathbf{x}^l_i\|$, where $\|\cdot\|$ represents the Euclidean distance. This maximum distance is plotted on the x-axis in Figure~\ref{fig:max_dist}, with the RMSD score between the corresponding predicted ligand pose and ground truth ligand plotted on the y-axis. The analysis shows that cases with larger maximum distances tend to exhibit larger RMSD errors. And the fixed radius of 20{\AA} in FABind is insufficient to encompass all the potential ligand structures.
Therefore, we are motivated to propose our solution: predicting a dynamic pocket radius to cover all atoms of the ligand as much as possible.

\noindent\textbf{Dynamic Pocket Radius Prediction.}
\label{sec:dy_pocket_r_pred}
To enhance pocket coverage for the entire ligand, we propose a dynamic pocket radius prediction module. Ideally, even if the pocket center prediction is incorrect, the enlarged pocket radius can still encompass most ligand atoms. 
To achieve this, we introduce a regression head for pocket radius prediction. The labels of the training data are set as the radii of the ground truth ligands, which hence can be viewed as a prediction of the ligand size\footnote{We have also tried other options as training labels, such as the distance between the predicted pocket center and the ground truth pocket center plus the ligand radius. In practice, we find that the ligand radius as the label is the best for stable training.}. Based on the predicted ligand size, we add an additional buffer to account for surrounding context interactions.

Denote the radius of the ground truth ligand conformation as $r$, with the updated hidden states $\mathbf{h}_i$ for ligand atoms from the pocket prediction module. The radius regression head, represented as $\phi_r$, employs a multilayer perceptron (\texttt{MLP}). The goal is to minimize the Huber loss~\cite{huber1992robust}:
\begin{equation}
L_r = \text{Huber}(r, \hat{r}), \quad 
\hat{r} = \phi_r(\sum_i{\mathbf{h}_i}).
\end{equation}
Then, similar to FABind, we calculate the predicted center of the classified residues from the pocket prediction module 
as the predicted pocket center. The pocket is then defined as a sphere around this predicted center with the predicted radius $\hat{r}$ plus a buffer $\beta$, $\hat{r} + \beta$, instead of a fixed radius.
In this way, we are able to cover most 
atoms in the ligand for the further docking part.

\subsubsection{Docking Structure Prediction}
After pocket prediction, FABind utilizes a scheduled sampling~\cite{bengio2015scheduled} training strategy that gradually incorporates the predicted pocket for docking training. We replace this approach with teacher forcing, using the ground-truth pocket center for docking training while still applying pocket center noise for generalization. This modification stabilizes the training process. Apart from this, within the docking module, we explore various ways to enhance docking performance.

\noindent\textbf{Permutation Loss.}
The rationality of molecular conformation is crucial for the performance of docking procedures. 
To improve the rationality of the generated conformations and reduce dependence on post-optimization, we introduce permutation loss~\cite{dmcg,zhang2022ligpose} into the docking model of \method{}. This loss is designed to ensure permutation invariance of symmetric atoms in molecular conformations during training. For example, as illustrated in the top-left part of Figure~\ref{fig:overview}, exchanging atom $x$ with $y$ or atom $a$ with $b$ is equivalent, yielding the same conformation when their coordinates are swapped. Therefore, we update the model using the lower loss resulting from these permutations. 
Denote the predicted ligand conformation as $\mathcal{\hat{R}}^l$ and the ground truth ligand conformation as $\mathcal{R}^l$. The permutation loss is defined as:
\begin{equation}
L_p = \min\limits_{\sigma \in \mathcal{S}}\{\text{Huber}(\mathcal{R}^l, \sigma(\mathcal{\hat{R}}^l))\},
\end{equation}
where $\mathcal{S}$ represents the set of permutation operations applied to symmetric atoms of the molecule. In practice, we use the graph tool\footnote{\url{https://graph-tool.skewed.de}} to extract all permutations of a ligand graph.

\subsection{From Regression to Sampling}
Sampling capability is necessary for capturing multiple binding sites and conformations. Previous sampling models typically involve training a generative model for this purpose~\cite{corso2023diffdock,wang2023flexidock,guo2023diffdock}. However, \method{} demonstrates that it is not necessary to train a generative model; instead, simple techniques can be applied to enable a regression model to achieve high-quality sampling capability, with a novel light confidence model to select the final pose.


\subsubsection{Sampling Method}

In DiffDock~\cite{corso2023diffdock}, the authors discussed the limitations of the regression model due to the possibility of different pockets and conformations in a target protein. Hence, in our sampling-based \method{}, we utilize a clustering method for pocket variant prediction. Besides, we adopt a simple dropout-based conformation generation method (dropout sampling) to produce conformation variants. Notably, our modified \method{} for the sampling version does not require training for pocket and conformation variant generation, making it lightweight and based on the regression-based \method{}.

\noindent\textbf{Clustering for Pocket Variants.}
\label{sec:cluster_pocket_var}
The pocket prediction module of FABind is designed to predict a single pocket center, which is insufficient for identifying multiple binding sites. For pocket variants, we adopt the method from \cite{braun2022mapping,yan2023multi} to use DBSCAN for pocket clustering. This algorithm does not require the pre-specification of cluster centers and exhibits a tolerance for noise, making it a versatile choice among clustering algorithms. During each sampling process, there is a probability $p$ (with $p=0.5$) that we randomly choose a cluster from the DBSCAN output; otherwise, the initially predicted pocket center is used. The center of this cluster is then adopted as the pocket center. This strategy aims at maximizing the diversity. Regardless of the method used to determine the pocket center, the radius predicted by the radius prediction module is employed to redefine the pocket boundaries in a spherical manner, ensuring sufficient coverage of potential ligand positions.

\noindent\textbf{Dropout Sampling for Conformation Generation.}
To sample variants of the ligand conformation, we use a very simple trick to leverage the randomness of dropout. Dropout~\cite{srivastava2014dropout} is a commonly adopted method to overcome overfitting and improve generalization. It randomly drops several units in each model layer and leads to a sub-model over the full model for forward passing. In this work, we utilize the property of randomness from dropout to do the sampling. 
During training, each sub-model produced by dropout is supervised by the ground truth conformation (label). As a result, the conformations generated by these sub-models are generally reasonable and close to the correct pose. Thus, we use each forward pass of the sub-model with dropout to generate the conformation variants.
The implementation is also easy, similar to R-Drop~\cite{wu2021r}, when generating a size of $s$ conformations for an input ligand $\textbf{x}$, we can simply repeat the input $\textbf{x}$ for $s$ times in a batch and go through one forward pass to generate $s$ conformations.

\begin{figure}
\centering
\includegraphics[width=\linewidth]{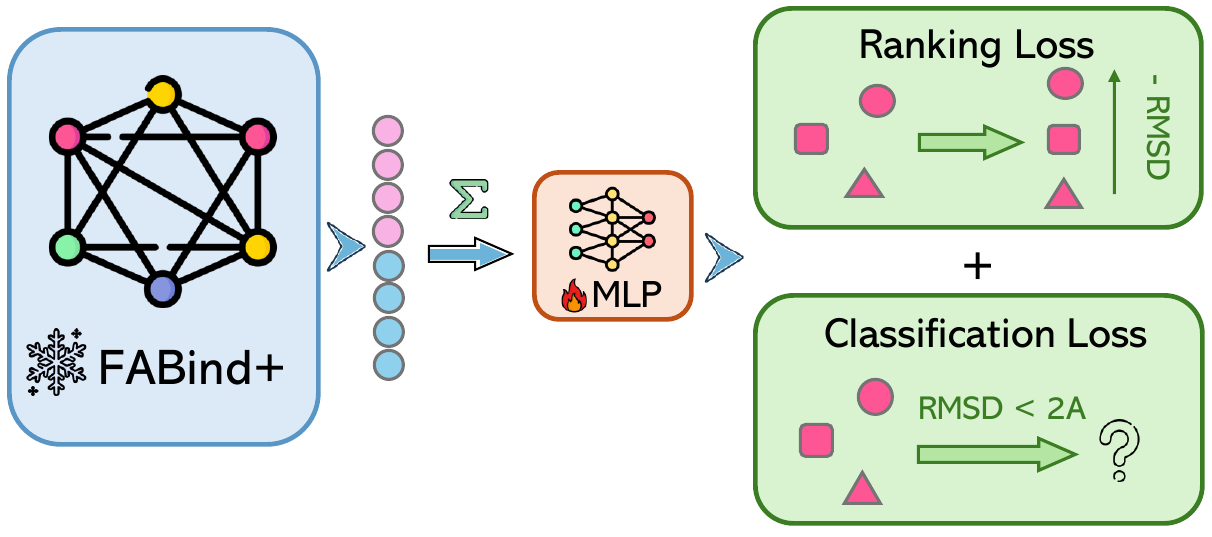}
\caption{Confidence model training pipeline. We add lightweight \texttt{MLP} layers upon the fixed FABind+ model for the confidence model, which incorporates a ranking loss and a classification loss for training the confidence model.}
\label{fig:confidence_model}
\end{figure}

\subsubsection{Confidence Model}
After generating multiple conformations as candidates, a confidence model is required to select the final pose. 
In this work, we introduce a lightweight confidence model comprising only several \texttt{MLP} layers following the sampling-based \method{} backbonew. On one hand, this approach is advantageous over previous work~\cite{corso2023diffdock} as it allows for on-the-fly collection of training data. On the other hand, more complex model architectures do not necessarily yield better performance, as illustrated in Appendix~\ref{supp:config}. The overall pipeline is shown in Figure~\ref{fig:confidence_model}.
During the training of the sampling model, with the parameters of \method{} frozen, we use pocket clustering and dropout sampling to explore different binding sites and conformations. The output states $\mathbf{h}_i$ from FABind+ are summed and then fed into the \texttt{MLP} layers to obtain the score, denoted as $s=f_\theta(\sum \mathbf{h}_i)$. Here, $f_\theta$ represents the confidence model, $\mathbf{h}_i$ is the output embedding from \method{}, and the scalar output $s$ is utilized for confidence loss computation.

\noindent\textbf{Training Objective.}
The training objective includes both classification loss, denoted as $L_{cls}$, and ranking loss, denoted as $L_{rank}$. The classification loss is formulated as a binary cross-entropy loss to accurately predict whether each pose has an RMSD below 2Å, following DiffDock~\cite{corso2023diffdock}. The motivation behind incorporating ranking loss is that achieving precise 2Å classification is challenging, whereas ranking between samples is comparatively simpler. Specifically, we adopt a pairwise ranking loss inspired by InstructGPT~\cite{ouyang2022training}. For each batch, each input ligand is repeated $N$ times, and there are then $N$ conformations generated after model forwarding, resulting $\binom{N}{2}$ for each protein-ligand complex. We train on all $\binom{N}{2}$ comparisons from each complex as a single batch element. According to InstructGPT, this method is computationally more efficient and reduces the likelihood of overfitting. The ranking loss is formally defined as follows:
\begin{equation}
L_{rank}=-\frac{1}{\binom{N}{2}} \mathbb{E}_{\left(\mathcal{R}_+, \mathcal{R}_-\right) \sim D}\left[\log \left(\sigma\left(s_+ - s_-\right)\right)\right],    
\end{equation}
where $s$ is the scalar score predicted by the confidence model for conformation $\mathcal{R}$, $\mathcal{R}_+$ is the preferred conformation with lower RMSD compared to $\mathcal{R}_-$, and $D$ is the dataset of comparison sets.

In the implementation, each GPU batch consists of one input ligand, which is then duplicated $N=5$ times. Following model forward propagation, each sample yields $N$ predicted structures. These structures are subsequently ranked according to their computed RMSD values, which are also utilized to calculate the classification loss $L_{cls}$. The total loss for the confidence model, denoted as $L_{conf}$, is the sum of the classification loss and the ranking loss:
\begin{equation}
L_{conf} = L_{cls} + L_{rank}.
\end{equation}

\section{Experiments}
\subsection{Experimental Setup}

\begin{table*}[t]
    \caption{Performance of flexible blind self-docking. The number of poses that DiffDock and \method{} sample is specified in parentheses. \method{} without parentheses denotes regression-based performance. The results of the sampling-based \method{} are averaged over three inference runs. Methods operating solely on CPU are marked with "*". All baseline results (with the exception of those from DiffDock, as reported in their paper) are sourced from \citet{pei2023fabind}. The best results are highlighted in \textbf{bold}, and the second-best scores are marked with an \underline{underline}.
    }
    \vspace{-0.1cm}
    \centering
    \scalebox{0.85}{
    \begin{tabular}{lcccccc|cccccc|c}
    \toprule
        & \multicolumn{6}{c}{\bf{Ligand RMSD}} & \multicolumn{6}{c}{\bf{Centroid Distance}} & \\
        \cmidrule(lr){2-7} \cmidrule(lr){8-13}
        & \multicolumn{4}{c}{Percentiles $\downarrow$} & \multicolumn{2}{c}{\% Below $\uparrow$} & \multicolumn{4}{c}{Percentiles $\downarrow$} & \multicolumn{2}{c}{\% Below $\uparrow$} & \bf{Average} \\
        \cmidrule(lr){2-5} \cmidrule(lr){6-7} \cmidrule(lr){8-11} \cmidrule(lr){12-13}
        \textbf{Method} & 25\% & 50\% & 75\% & Mean & 2\AA & 5\AA & 25\% & 50\% & 75\% & Mean & 2\AA & 5\AA & \bf{Runtime (s)}\\
    \midrule
        \rowcolor[RGB]{234, 238, 234} \multicolumn{14}{l}{\textit{Traditional docking software}} \\
        \textsc{QVina-W} & 2.5 & ~~7.7 & 23.7 & 13.6 & 20.9 & 40.2 & 0.9 & 3.7 & 22.9 & 11.9 & 41.0 & 54.6 & ~~~~49*\\
        \textsc{GNINA} & 2.8 & ~~8.7 & 22.1 & 13.3 & 21.2 & 37.1 & 1.0 & 4.5 & 21.2 & 11.5 & 36.0 & 52.0 & 146\\
        \textsc{SMINA} & 3.8 & ~~8.1 & 17.9 & 12.1 & 13.5 & 33.9 & 1.3 & 3.7 & 16.2 & ~~9.8 & 38.0 & 55.9 & ~~146*\\
        \textsc{GLIDE} & 2.6 & ~~9.3 & 28.1 & 16.2 & 21.8 & 33.6 & 0.8 & 5.6 & 26.9 & 14.4 & 36.1 & 48.7 & ~1405*\\
        \textsc{Vina} & 5.7 & 10.7 & 21.4 & 14.7 & ~~5.5 & 21.2 & 1.9 & 6.2 & 20.1 & 12.1 & 26.5 & 47.1 & ~~205*\\
    \midrule
        \rowcolor[RGB]{234, 238, 234} \multicolumn{14}{l}{\textit{Deep learning-based method}} \\
        \textsc{EquiBind} & 3.8 & ~~6.2 & 10.3 & 8.2 & ~~5.5 & 39.1 & 1.3 & 2.6 & ~~7.4 & 5.6 & 40.0 & 67.5 & ~~\textbf{0.03}\\
        \textsc{TankBind} & 2.6 & ~~4.2 & ~~7.6 & 7.8 & 17.6 & 57.8 & 0.8 & 1.7  & ~~4.3 & 5.9 & 55.0  & 77.8 & ~~0.87\\
        \textsc{E3Bind} & 2.1 & ~~3.8 & ~~7.8 & 7.2 & 23.4 & 60.0 & 0.8 & 1.5 & ~~4.0 & 5.1 & 60.0 & 78.8 & ~~0.44\\
        \textsc{DiffDock} (10) & 1.5 & ~~3.6 & ~~7.1 & - & 35.0 & 61.7 & \underline{0.5} & 1.2 & ~~3.3 & - & 63.1 & 80.7 & 20.81\\
        \textsc{DiffDock} (40) & 1.4 & ~~3.3 & ~~7.3 & - & 38.2 & 63.2 & \underline{0.5} & 1.2 & ~~3.2 & - & 64.5 & 80.5 & 82.83\\
        \textsc{FABind} & 1.7 & ~~3.1 & ~~6.7 & \underline{6.4} & 33.1 & 64.2 & 0.7 & 1.3 & ~~3.6 & \underline{4.7} & 60.3 & 80.2 & ~~\underline{0.12} \\
    \midrule
        \rowcolor[RGB]{234, 238, 234} \multicolumn{14}{l}{\textit{Our model}} \\
        \textsc{Regression \method{}} & \textbf{1.2} & ~~\underline{2.6} & ~~5.8 & \textbf{5.2} & \underline{43.5} & 71.1 & \textbf{0.4} & \textbf{1.0} & 2.9 & ~~\textbf{3.5} & 67.5 & 84.0 & ~~0.16 \\
        \textsc{Sampling \method{} (10)} & \underline{1.3} & ~~2.7 & ~~\textbf{5.4} & \textbf{5.2} & 42.4 & \textbf{71.6} & \underline{0.5} & \underline{1.1} & \underline{2.8} & ~~\textbf{3.5} & \underline{67.8} & \underline{84.6} & ~1.6 \\
        \textsc{Sampling \method{} (40)} & \textbf{1.2} & ~~\textbf{2.4} & ~~\underline{5.6} & \textbf{5.2} & \textbf{44.9} & \underline{71.3} & \underline{0.5} & \textbf{1.0} & \textbf{2.7} & ~~\textbf{3.5} & \textbf{68.3} & \textbf{85.2} & ~6.4 \\
        
    \bottomrule
    \end{tabular}
    }
    \label{tab:blind_self_docking}
\end{table*}

\textbf{Dataset.} We evaluate our methods on PDBbind v2020 dataset~\cite{liu2017forging}, curated from Protein Data Bank (PDB)~\cite{burley2021rcsb}. Consistent with the data split method of EquiBind, as adopted in most prior research~\cite{corso2023diffdock, stark2022equibind, lu2022tankbind, pei2023fabind}, we used structures published before 2019 for training and those from 2019 onwards for testing. We excluded proteins with over 1500 residues and ligands with more than 150 atoms, resulting in 17,644 samples for training, 958 for validation, and 363 for testing. Further details about our preprocessing steps are outlined in Appendix~\ref{supp:data}.

\noindent\textbf{Baselines.} \method{} is benchmarked against many traditional methods and deep learning models. 
QVINA-W, GNINA~\cite{mcnutt2021gnina}, SMINA~\cite{koes2013lessons}, GLIDE~\cite{friesner2004glide} and VINA~\cite{trott2010autodock} are concluded as the powerful traditional methods. 
Deep learning models can be categorized into two main classes: sampling-based and regression-based models. For the sampling-based models, we include DiffDock~\cite{corso2023diffdock} with different sample size, 
and for regression-based models, we include EquiBind~\cite{stark2022equibind} TankBind~\cite{lu2022tankbind}, E3Bind~\cite{zhang2022e3bind}, and FABind~\cite{pei2023fabind}.

\noindent\textbf{Evaluation Metrics.} Our evaluation employs two key metrics: (1) \textit{Ligand RMSD}, which measures the root-mean-square deviation (RMSD) between the predicted and ground-truth ligand coordinates. We report the symmetry-corrected RMSD using sPyRMSD\footnote{\url{https://github.com/RMeli/spyrmsd}}~\cite{meli2020spyrmsd}. Detailed descriptions can be found in Appendix~\ref{supp:metric}. (2) \textit{Centroid Distance}, which calculates the Euclidean distance between the centroids of the predicted and true ligand structures. 

\noindent\textbf{Implementation Details.}
We recognize that the capability of the docking module is essential for achieving optimal results. Ideally, larger models would trade speed for higher accuracy. However, as shown in Figure~\ref{fig:ablation_model_size}, we observe that larger models struggle with optimization. Based on this observation, we decided to simply extend the model by incorporating an additional FABind layer. A detailed analysis of this approach, along with other design techniques, is provided in Appendix~\ref{supp:model}.

\subsection{Main Results}

\subsubsection{Blind Self-Docking Performance.}
Blind self-docking involves docking a flexible ligand to a protein without prior knowledge of the exact binding site, requiring accurate predictions of the translation, rotation, and conformation of the ligand. 
As shown in Table~\ref{tab:blind_self_docking}, our regression-based approach significantly outperforms the powerful generative model DiffDock. It achieves a success rate of 43.5\% for ligand atomic RMSD less than 2Å, surpassing DiffDock by 5.3 percentage points. This model demonstrates superior accuracy across the board, as evidenced by improvements in both the mean RMSD and the percentage of predictions under 2Å and 5Å.

\begin{figure}
\centering
\includegraphics[width=0.8\linewidth]{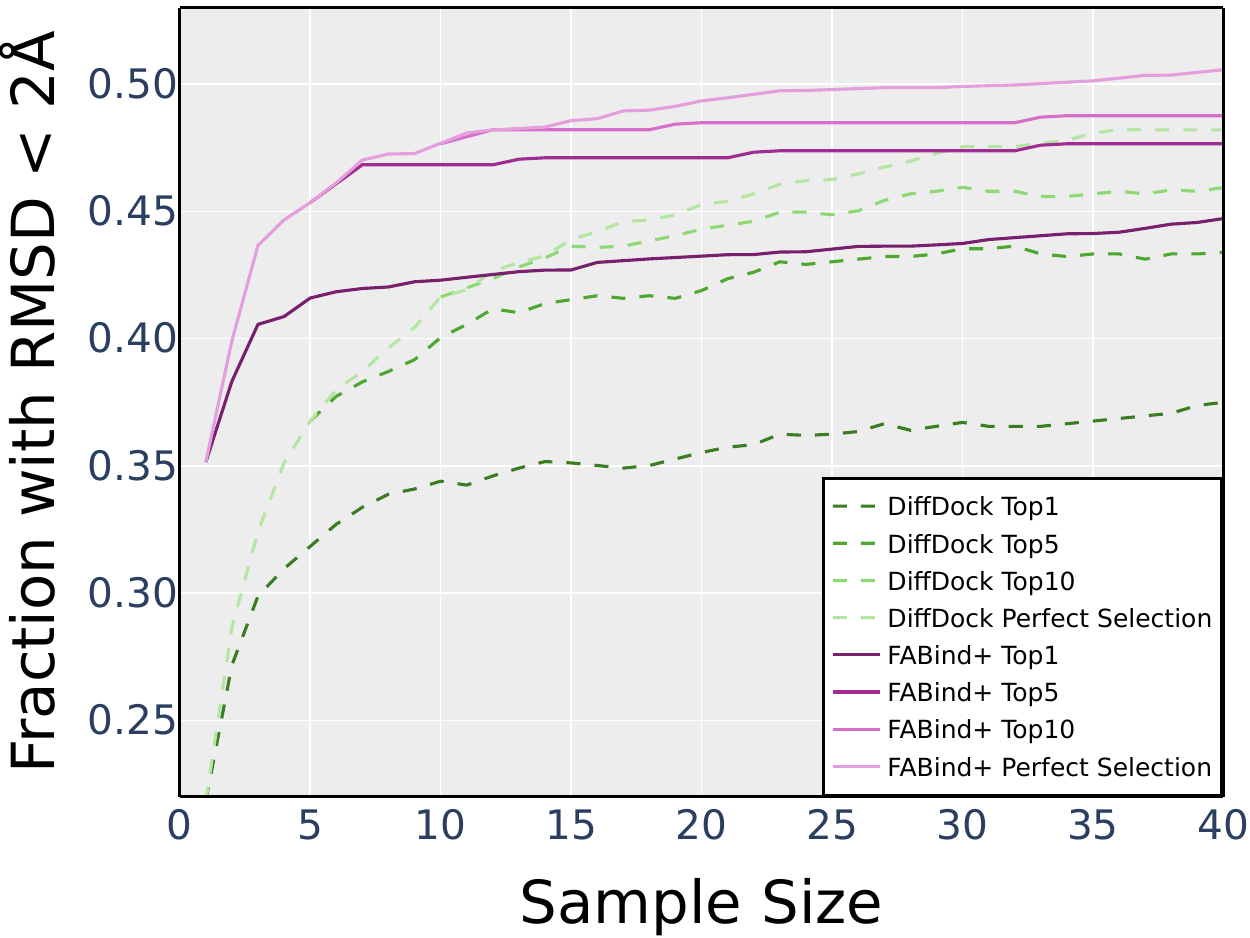}
\vspace{-0.2cm}
\caption{Scaling curve with increasing sample size. `` Perfect Selection'' refers to refers to choosing the samples with the lowest RMSD at each given sample size.}
\label{fig:sample_diversity}
\vspace{-0.1cm}
\end{figure}

On the other hand, our sampling-based model shows better results, especially as the sample size increases. As shown in Figure~\ref{fig:sample_diversity}, with a sample size of 1, we achieve 35.3\% of predictions under an RMSD of 2Å, which is reasonable since sampling models are less likely to produce highly accurate results with a single shot compared to the same predictive models. With a sample size of 10, the model slightly outperforms its regression-based version. As we expand the sample size to 40, similar to DiffDock's setting, our sampling-based model delivers superior performance across most metrics. \method{} with a sample size of 40 achieves a performance of 44.9\%. In addition to accurate predictions, we maintain a fast sampling speed compared to other sampling-based models. The inference speed of the sampling-based \method{} is 13 times faster than DiffDock under equivalent conditions.
Notably, while the performance gains from the sampling version may not appear significant in the table, its strength lies in sampling multiple pockets and generating diverse conformations, which are crucial abilities that the regression model cannot achieve.

\subsubsection{Blind Self-Docking Performance for Unseen Proteins.} 
Here we seek to assess the generalization capability of \method{} on proteins not encountered during the training phase. 
Following previous works~\cite{zhang2022e3bind,pei2023fabind}, we evaluate the performance of \method{} on a set of proteins filtered based on their UniProt IDs, specifically retaining only samples whose proteins are unseen in the training and validation stages. The results of this evaluation are summarized in Table~\ref{tab:blind_self_docking_unseen}. It is evident from the findings that \method{} exhibits superior performance, surpassing all baseline methods and its predecessor, FABind, across all metrics. 
This performance underscores the effectiveness of our enhanced design in achieving robust generalization on unseen proteins.

\begin{table*}
    \caption{Performance of flexible blind self-docking on unseen receptors. Sampling-based \method{} results are averaged over three inference runs. All baseline results are sourced from \citet{pei2023fabind}.}
    \centering
    \vspace{-0.2cm}
    \scalebox{0.9}{
    \begin{tabular}{lcccccc|cccccc}
    \toprule
        & \multicolumn{6}{c}{\bf{Ligand RMSD}} & \multicolumn{6}{c}{\bf{Centroid Distance}} \\
        \cmidrule(lr){2-7} \cmidrule(lr){8-13}
        & \multicolumn{4}{c}{Percentiles $\downarrow$} & \multicolumn{2}{c}{\% Below $\uparrow$} & \multicolumn{4}{c}{Percentiles $\downarrow$} & \multicolumn{2}{c}{\% Below $\uparrow$} \\
        \cmidrule(lr){2-5} \cmidrule(lr){6-7} \cmidrule(lr){8-11} \cmidrule(lr){12-13}
        \textbf{Method} & 25\% & 50\% & 75\% & Mean & 2\AA & 5\AA & 25\% & 50\% & 75\% & Mean & 2\AA & 5\AA \\
    \midrule
        \rowcolor[RGB]{234, 238, 234} \multicolumn{13}{l}{\textit{Traditional docking software}} \\
        \textsc{QVina-W} & 3.4 & 10.3 & 28.1 & 16.9 & 15.3 & 31.9 & 1.3 & ~~6.5 & 26.8 & 15.2 & 35.4 & 47.9 \\
        \textsc{GNINA} & 4.5 & 13.4 & 27.8 & 16.7 & 13.9 & 27.8 & 2.0 & 10.1 & 27.0 & 15.1 & 25.7 & 39.5 \\
        \textsc{SMINA} & 4.8 & 10.9 & 26.0 & 15.7 & ~~9.0 & 25.7 & 1.6 & ~~6.5 & 25.7 & 13.6 & 29.9 & 41.7 \\
        \textsc{GLIDE} & 3.4 & 18.0 & 31.4 & 19.6 & 19.6 & 28.7 & 1.1 & 17.6 & 29.1 & 18.1 & 29.4 & 40.6 \\
        \textsc{Vina} & 7.9 & 16.6 & 27.1 & 18.7 & ~~1.4 & 12.0 & 2.4 & 15.7 & 26.2 & 16.1 & 20.4 & 37.3 \\
    \midrule
        \rowcolor[RGB]{234, 238, 234} \multicolumn{13}{l}{\textit{Deep learning-based method}} \\
        \textsc{EquiBind} & 5.9 & 9.1 & 14.3 & 11.3 & ~~0.7 & 18.8 & 2.6 & 6.3 & 12.9 & 8.9 & 16.7 & 43.8 \\
        \textsc{TankBind} & 3.4 & 5.7 & 10.8 & 10.5 & ~~3.5 & 43.7 & 1.2 & 2.6 & ~~8.4 & 8.2 & 40.9 & 70.8 \\
        \textsc{E3Bind} & 3.0 & 6.1 & 10.2 & 10.1 & ~~6.3 & 38.9 & 1.2 & 2.3 & ~~7.0 & 7.6 & 43.8 & 66.0 \\
        \textsc{DiffDock (10)} & 3.2 & 6.4 & 16.5 & 11.8 & 14.2 & 38.7 & 1.1 & 2.8 & 13.3 & 9.3 & 39.7 & 62.6 \\
        \textsc{DiffDock (40)} & 2.8 & 6.4 & 16.3 & 12.0 & 17.2 & 42.3 & 1.0 & 2.7 & 14.2 & 9.8 & 43.3 & 62.6 \\
        \textsc{FABind} & \underline{2.2} & 3.4 & ~~\textbf{8.3} & ~~7.7 & 19.4 & 60.4 & 0.9 & \underline{1.5} & ~~4.7 & 5.9 & 57.6 & 75.7 \\
    \midrule
        \rowcolor[RGB]{234, 238, 234} \multicolumn{13}{l}{\textit{Our model}} \\
        \textsc{Regression \method{}} & \textbf{1.6} & \underline{3.3} & ~~8.9 & ~~\textbf{7.0} & \underline{34.7} & \textbf{63.2} & \textbf{0.5} & \underline{1.5} & ~~\textbf{4.2} & \textbf{5.1} & \underline{58.3} & \textbf{77.1} \\
        \textsc{Sampling \method{}(10)} & \textbf{1.6} & \textbf{3.2} & ~~9.0 & ~~7.4 & 33.3 & \underline{61.8} & \underline{0.6} & \textbf{1.4} & ~~\underline{4.3} & 5.7 & \textbf{59.0} & 75.0 \\
        \textsc{Sampling \method{}(40)} & \textbf{1.6} & \underline{3.3} & ~~\underline{8.8} & ~~\underline{7.1} & \textbf{35.4} & 61.1 & \underline{0.6} & \underline{1.5} & ~~4.9 & \underline{5.3} & \underline{58.3} & \underline{76.3} \\
    \bottomrule
    \end{tabular}
    }
    \label{tab:blind_self_docking_unseen}
    
\end{table*}

\subsubsection{Performance Scaling with Increased Sample Size.} To enhance the understanding of the sampling capacity, we depict ``Top1'', ``Top5'', ``Top10'', and ``Perfect Selection'' across varying sample sizes in Figure~\ref{fig:sample_diversity}. DiffDock stands as a highly powerful generative model, and akin to DiffDock, our approach shows significant performance gains as the sample size increases. When the sample size reaches 40, we are able to dock 51.2\% of samples with an RMSD lower than 2Å. Our growth curves for the top1, top5, and top10 selections also mirror those of DiffDock. The observation of similar trends suggests that our lightweight confidence model is equally effective. In contrast to their 20M parameter model, our model is substantially smaller, with only 0.8M parameters, and based on a simple \texttt{MLP}. Indeed, while our curves outperform those of DiffDock, this advantage begins with our random selection (sample size = 1) outperforming DiffDock by about 15\%. This does not necessarily indicate that the sampling ability brought forth by dropout or pocket clustering surpasses that of diffusion or other generative methods. However, it does imply that achieving enhanced sampling capacity in docking does not necessarily require generative training.

\section{Analysis}
\subsection{Component Analysis}

\begin{figure}
\centering
\includegraphics[width=0.8\linewidth]{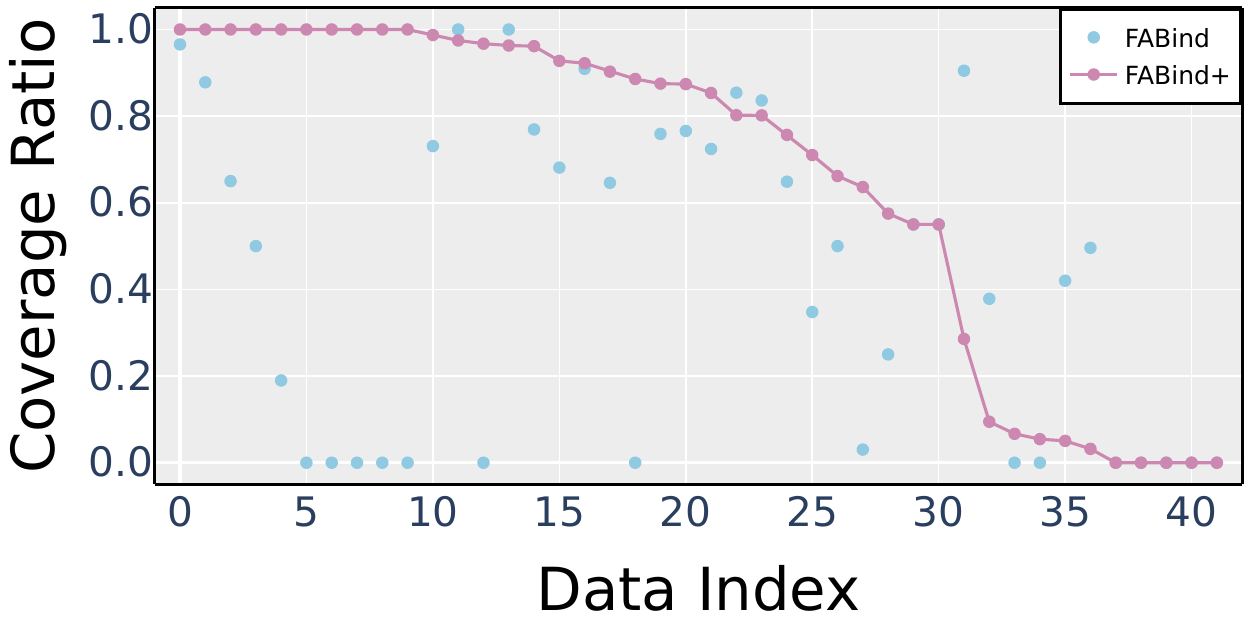}
\caption{Coverage ratio comparison between FABind (blue) and \method{} (pink). Samples are sorted in descending order according to the coverage ratio of \method{}, with points connected for clarity.}
\vspace{-0.4cm}
\label{fig:test_coverage}
\end{figure}

\noindent\textbf{Pocket Radius Module Analysis.}
As discussed in Section~\ref{sec:dy_pocket_r_pred}, we introduce dynamic pocket radius prediction to make the pocket cover all atoms of the ligand as much as possible.
Here we further compare the predicted pockets from FABind and \method{}. 
We focus on the coverage ratio performance within a subset of the test set, defined by the union of test samples where either FABind or \method{} could not achieve a 100\% coverage ratio. 
The results, as depicted in Figure~\ref{fig:test_coverage}, reveal that within this collective of test samples, \method{} demonstrates superior performance, with the coverage ratio for the majority of \method{} predictions being higher than that of FABind. 
Such outcomes demonstrate the effectiveness of our proposed dynamic pocket radius prediction.

\begin{figure}[htpb]
\centering
\includegraphics[width=0.8\linewidth]{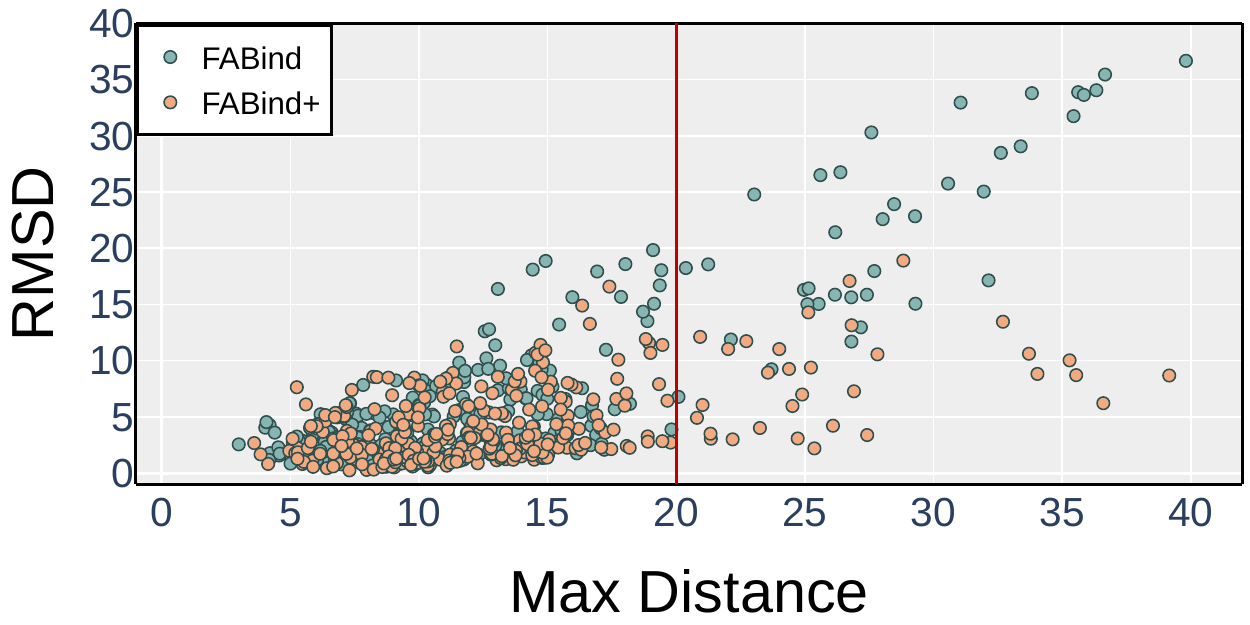}
\caption{Analysis of pocket prediction and the predicted ligand RMSD in the test set. ``Max Distance'' is the distance between the predicted pocket center and the farthest ground truth ligand atom.}
\label{fig:test_max_dist}
\vspace{-0.4cm}
\end{figure}

\begin{figure}
\centering
\includegraphics[width=0.8\linewidth]{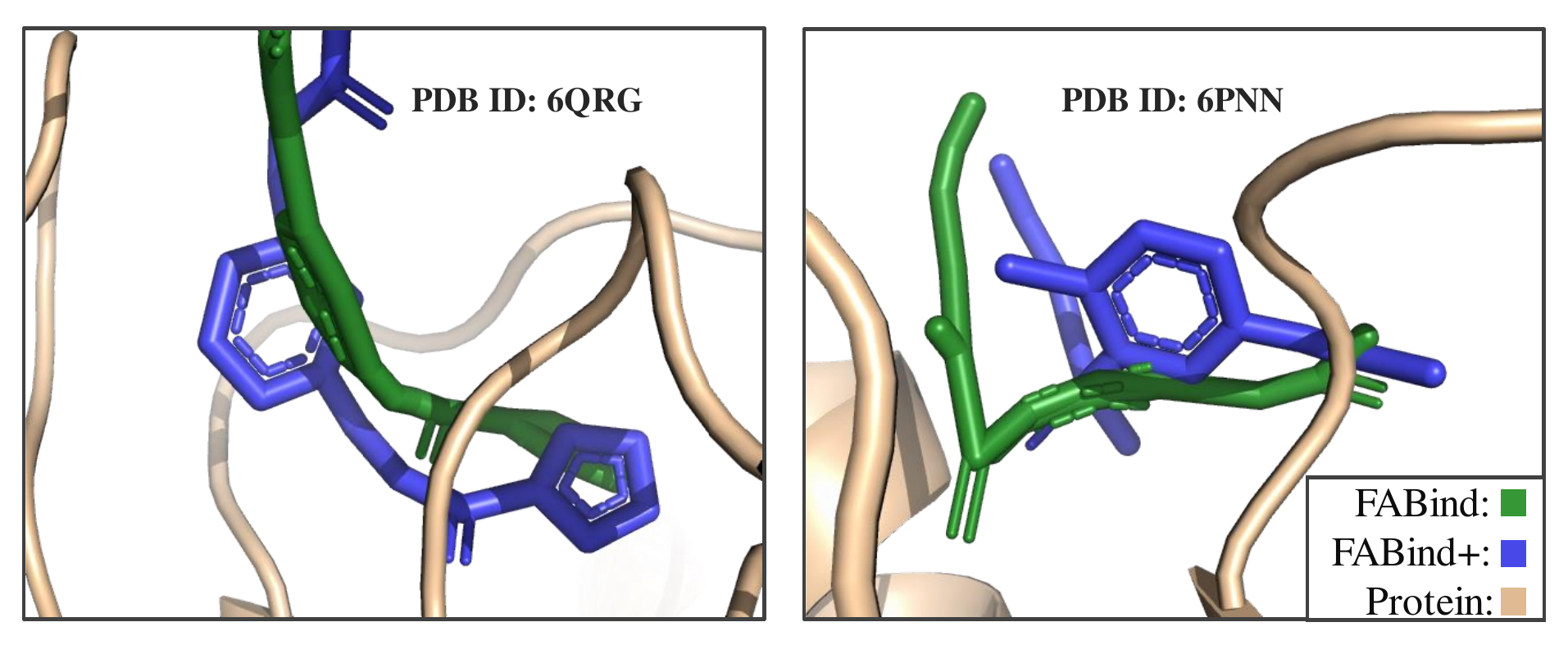}
\caption{Cases (PDB 6QRG and PDB 6PNN) for permutation-invariant loss analysis. We show the ground truth protein and the predicted conformations by FABind in green and \method{} in blue.}
\label{fig:permutation_loss_analy}
\end{figure}

In reference to Figure~\ref{fig:max_dist}, we further compare the pocket prediction performance between FABind and \method{} on the test set, as depicted in Figure~\ref{fig:test_max_dist}. ``Max Distance'' represents the distance between the predicted pocket center and the farthest 
ground truth ligand atom. This offers an approximation of the pocket size required to encompass the ligand based on the predicted pocket center. From this Figure~\ref{fig:test_max_dist}, we can see that \method{} shows better performance with a lower RMSD score than FABind across samples with different max distances, which demonstrates the effectiveness of our enhanced designs. For those samples with a large max distance, our method's precise pocket prediction and dynamic radius prediction capabilities enable more accurate prediction. 

\noindent\textbf{Permutation-invariant Loss Analysis.}
To elucidate the impact of the permutation-invariant loss on the generated conformations, we present two cases in Figure~\ref{fig:permutation_loss_analy} from FABind (without permutation-invariant loss) and \method{}, respectively.
Ligands in these cases contain rings that are locally symmetric. 
FABind fails to generate the conformations of these rings, with the atoms on the rings being aligned linearly, whereas \method{} successfully predicts their conformations. 
This discrepancy arises because the model encounters samples with the same symmetric substructure patterns, such as benzene rings, during training, though the ordering of the atoms may differ. 
Without the permutation-invariant loss, the model might learn a variety of local optima associated with different atom arrangements. 
These cases further underscore the significance of permutation-invariant loss in conformation prediction.

\begin{figure}
\centering
\includegraphics[width=0.7\linewidth]{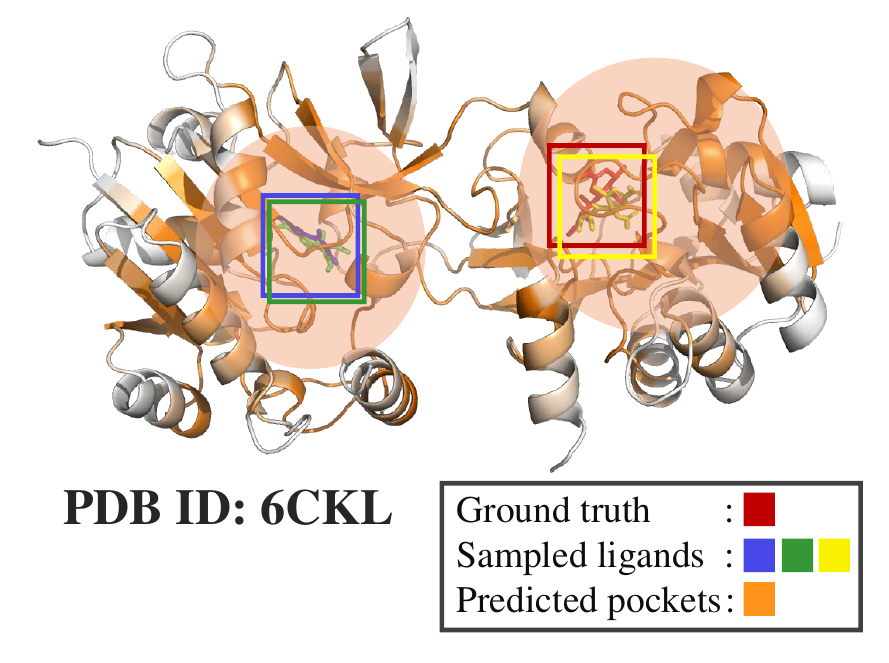}
\vspace{-0.1cm}
\caption{Case for pocket clustering.}
\vspace{-0.4cm}
\label{fig:pocket_cluster_analy}
\end{figure}

\noindent\textbf{Pocket Clustering Analysis.}
Here we conduct a case study to show the effectiveness of our pocket clustering method in detecting multiple binding sites, as discussed in Section~\ref{sec:cluster_pocket_var}.
We select the protein (PDB 6CKL), which is characterized by its symmetric arrangement of two chains. This symmetry results in two ground truth pockets along with their corresponding ligand conformations. 
As depicted in Figure~\ref{fig:pocket_cluster_analy}, with our clustering approach, \method{} successfully identifies both pockets and the corresponding docking conformations through sampling methods. 
In contrast, regression models are unable to capture this scenario. This case underscores the proficiency of the sampling-based \method{}.

\begin{figure}
\centering
\includegraphics[width=\linewidth]{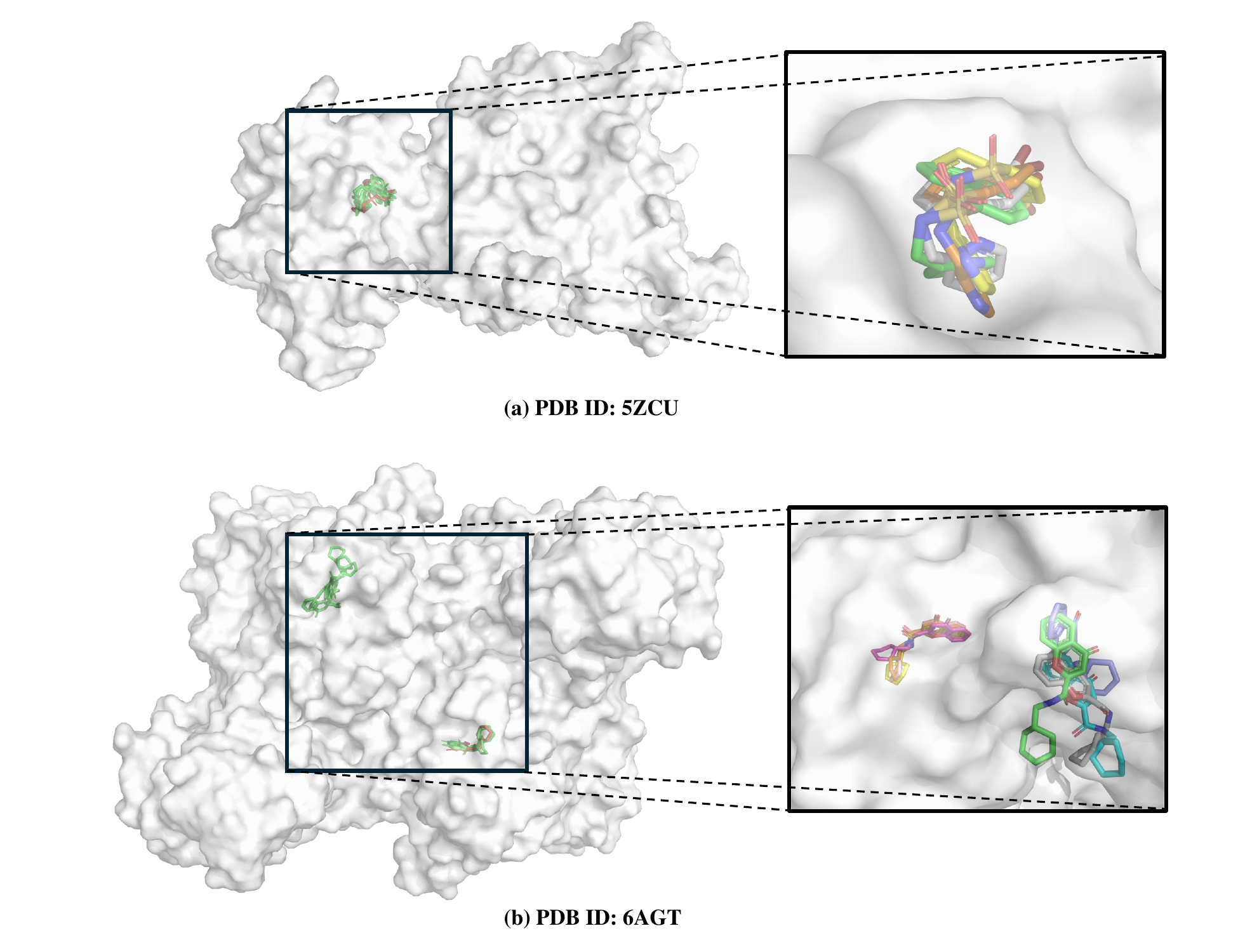}
\caption{Sampling diversity of \method{}. On the left, sampled structures (green) alongside the ground truth (red) are depicted; on the right, a closer view of the sampled structures is provided, each colored randomly for distinction.}
\label{fig:sam_showcase}
\end{figure}

\noindent\textbf{Visualizations of Sampling Capability.}  We also showcasing the sampling results in Figure~\ref{fig:sam_showcase}. In the case of complex 5ZCU, which features a single binding site, \method{} successfully generates a variety of conformations that are centered around the ground truth. For complex 6AGT, characterized by multiple pockets, we can observe that \method{} not only identifies these pockets but also generates diverse structures within each pocket.

\subsection{Ablation Study}

\begin{table}[htpb]
\caption{Ablation study on ligand RMSD metric.}
\vspace{-0.2cm}
\label{tab:ablation_results}
\begin{center}
\begin{small}
\begin{sc}
\begin{tabular}{l|ccc}
\toprule
Parameter & Below < 2Å $\uparrow$ & Mean$\downarrow$ & Med $\downarrow$ \\    
\midrule
regression \method{} & \textbf{43.5}\% & \textbf{5.2} & \textbf{2.6} \\
 - w/o. dynamic rad & 40.0\% & 5.5 & 2.8 \\
 - w/o. permutation loss & 41.2\% & 5.6 & 2.8 \\
 - w/o. add one layer & 39.1\% & 5.5 & 2.7 \\ 
\midrule
sampling \method{} (10) & \textbf{42.4}\% & \textbf{5.2} & \textbf{2.6} \\
 - w/o. confidence model & 35.3\% & 6.4 & 3.1 \\
 - classification loss only & 41.9\% & 5.4 & 2.6 \\
 - ranking loss only & 41.7\% & 5.5 & 2.7 \\
\bottomrule
\end{tabular}
\end{sc}
\end{small}
\end{center}
\end{table}

We conducted a detailed ablation study to evaluate the impact of various components on both the regression and sampling versions of FABind+. The results are summarized in the Table~\ref{tab:ablation_results}. 

For the regression version, removing the dynamic radius adjustment (`- w/o. dynamic rad`), permutation loss (`- w/o. permutation loss`), or additional layer (`- w/o. add one layer`) resulted in accuracy drops, with the success rate for predictions under 2Å decreasing from 43.5\% to 40.0\%, 41.2\%, and 39.1\%, respectively. The mean and median RMSD also worsened without these design elements.

In the sampling version (with a sample size of 10), the absence of the confidence model 
(`- w/o. confidence mode`) caused a significant drop in performance, reducing the success rate from 42.4\% to 35.3\%. We also compared the effects of using only classification loss or only ranking loss. The results showed that while both losses contribute to performance, using them together yields the best results.

\begin{figure}
\centering
\includegraphics[width=0.8\linewidth]{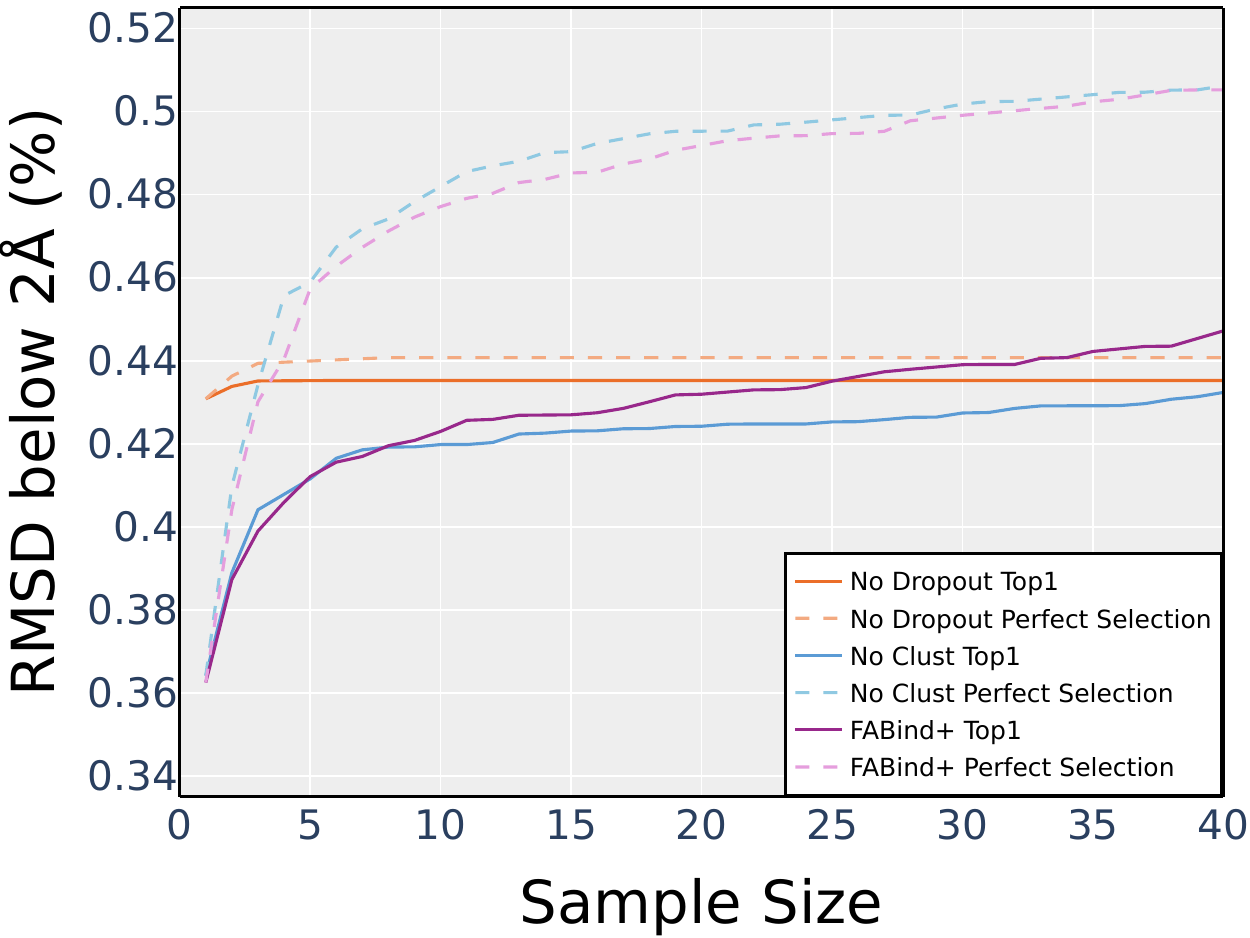}
\caption{Ablation analysis for the sampling version.}
\label{fig:ablation_sampling}
\end{figure}

\noindent\textbf{Ablations for Sampling Capability.} 
For the sampling version, we additionally present a scaling curve in Figure~\ref{fig:ablation_sampling} to illustrates how performance changes for different ablations.
This figure underscores the importance of each design element in achieving optimal performance. Specifically, removing pocket clustering (`No Clust`) leads to a slight drop in performance, as it hinders the ability to capture multiple binding sites. The lack of dropout sampling (`No Dropout`) means that diversity relies entirely on pocket clustering. However, given the limited number of pockets, the orange dashed line in the graph exhibits only a minimal increase with sample size.

\section{Conclusion}
\vspace{-0.3em}
In this work, we demonstrate that \method{} can serve as a unified model for both regression and sampling. We enhance the original FABind method by introducing dynamic pocket radius prediction and permutation loss. Additionally, we transform the pre-trained regression model into a sampling model using pocket clustering and dropout sampling without further training. This approach to activating sampling capability is universal to all regression-based models. A lightweight confidence model is then trained to select the best conformation. Our results show that \method{} significantly outperforms FABind, achieving superior docking performance.
\begin{acks}
This work is supported by National Natural Science Foundation (62076105).
\end{acks}
\bibliographystyle{ACM-Reference-Format}
\balance
\bibliography{reference}

\appendix
\section{Experiment Details}
\subsection{Dataset Processing}
\label{supp:data}
We keep the dataset split consistent with previous works~\cite{lu2022tankbind, guo2023diffdock, pei2023fabind}. The selected 968 validation structures are from data before 2019, and 363 test structures are from 2019 onwards, both ensuring no ligand overlap with the training set. For the training set, we filter out a few samples that could not be read by RDKit or TorchDrug\footnote{https://github.com/DeepGraphLearning/torchdrug}, leaving 17,795 complexes. We retain protein chains with the nearest atom of the small molecule within 10 Å. 
Subsequently, samples with protein chains longer than 1500 amino acids, molecules larger than 150 atoms, or inadequate contact (less than 5 amino acids within 10 Å of molecule atoms) were excluded. These criteria resulted in the exclusion of 118, 32, and 1 sample, successively, ultimately yielding 17,644 samples for training.

\subsection{Symmetry-Corrected RMSD}
\label{supp:metric}
In the reported results, the root-mean-square deviation (RMSD) is calculated using sPyRMSD~\cite{meli2020spyrmsd}. sPyRMSD utilizes a graph matching tool to identify all possible graph isomorphisms, returning the minimum RMSD. This process is aligned with our permutation loss computation. This approach is consistent with our computation of permutation loss. In practice, given the limited number of isomorphic molecules in the test set, we observe a negligible performance gain compared to standard RMSD calculations.

\subsection{Permutation Loss Implementation}
\begin{table}[htpb]
    \centering
    \caption{Statistics on the number of permutations.}
    \scalebox{0.8}{
        \begin{tabular}{l|ccccccc}
        \toprule
        No. Permutation &  $\leq 2$ & $\leq 4$ & $\leq 8$ & $\leq 16$ & $\leq 32$ & $\leq 64$ & $\leq 128$ \\
        \midrule
        Percentage & 58.8\% & 75.9\% & 87.7\% & 94.5\% & 97.7\% & 98.8\% & 99.4\% \\
        \bottomrule
        \end{tabular}
    }
    \label{tab:permu_stat}
\end{table}

We utilize the graph-tool toolkit\footnote{https://graph-tool.skewed.de/} to identify all symmetric atoms following previous work~\cite{dmcg}. According to sPyRMSD~\cite{meli2020spyrmsd}, both graph-tool and networkx\footnote{https://networkx.org} are capable of extracting all permutations. Specifically, an isomorphism between graphs A and B is a bijective mapping of the vertices of graph A to vertices of graph B that preserves the edge structure of the graphs (molecular connectivity in the case of molecular graphs). The problem of finding symmetric atoms can be converted to a graph isomorphism problem.

The implementation of graph-tool is based on VF2 algorithm~\cite{cordella2004sub}. In \method{}, the identification of symmetric atoms is performed offline with multi-processing. This preprocessing takes a few hours to extract all possible permutations of the PDBBind dataset, which is relatively short compared to the training time. Also, it will not increase training cost since the average number of permutations on our used dataset is only $10.1$. The detailed statistics are in Table~\ref{tab:permu_stat}.

\section{Model Details of \method{}}
\label{supp:model}

\noindent\textbf{MLP Configuration.}
All multilayer perceptrons (MLPs) implemented in \method{} consist of a Layer Normalization, followed by two linear transformations with ReLU activations. An additional ReLU activation is applied after the final linear transformation if the output of the MLP is an embedding. To regulate the parameters of \method{}, the MLP hidden scale is set to 1, indicating that the MLP's hidden size is the same as the input embedding dimension. On the other hand, in the confidence model, the MLP hidden scale is increased to 4 to maximize the model capacity.\noindent\textbf{MLP Configuration.}
All multilayer perceptrons (MLPs) implemented in \method{} consist of a Layer Normalization, followed by two linear transformations with ReLU activations. An additional ReLU activation is applied after the final linear transformation if the output of the MLP is an embedding. To regulate the parameters of \method{}, the MLP hidden scale is set to 1, indicating that the MLP's hidden size is the same as the input embedding dimension. On the other hand, in the confidence model, the MLP hidden scale is increased to 4 to maximize the model capacity.

\begin{figure}[htpb]
\centering
\includegraphics[width=0.7\linewidth]{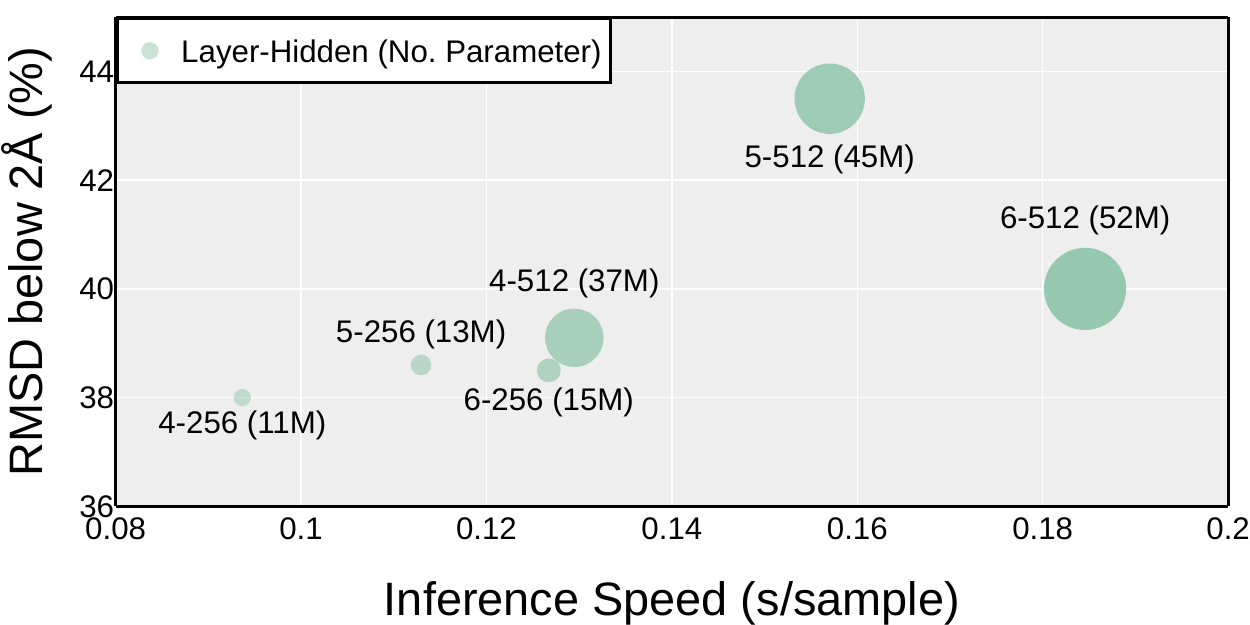}
\caption{Exploration of model size.}
\label{fig:ablation_model_size}
\end{figure}

\noindent\textbf{Model Size and Inference Speed.} 
We conduct comprehensive studies on the relationship between model size, performance, and inference speed, as illustrated in Figure~\ref{fig:ablation_model_size}. The x-axis denotes inference speed, while the y-axis represents performance, specifically the fraction of RMSD below 2Å. The size of the circles in the figure correlates with the number of parameters in the model. This figure suggests that larger models typically offer better performance but at the expense of reduced inference speed. Notably, the largest model setting does not yield improved results, potentially due to the need for meticulous hyperparameter tuning to unlock optimal performance.

\noindent\textbf{Implementation Details.}
Initial ligand conformations are generated using the ETKDG algorithm~\cite{riniker2015better}, followed by MMFF optimization via RDKit~\cite{landrum2013rdkit}, which produces the random generation of a low-energy ligand conformation. For pocket radius prediction, considering the fixed radius of FABind is 20Å, we adjust any predicted radius falling below this 20Å threshold to 20Å. This adjustment ensures the predicted pocket is at least as large as those predicted by FABind. For other hyperparameters and training configurations, we put the details in Appendix~\ref{supp:config}.

\section{Training and Inference}
\label{supp:config}

\begin{table*}[htpb]
\caption[SearchSpace]{The hyperparameter options we searched through for \method{}. The final parameters are marked in \textbf{bold}.}
\label{tab:fabind_hyperparameters}
\begin{center}
\begin{small}
\begin{sc}
\begin{tabular}{lc}
\toprule
Parameter & Search Space  \\    
\midrule
\rowcolor[RGB]{234, 238, 234} \multicolumn{2}{l}{\textit{Model Config}} \\
radius prediction from & protein, \textbf{ligand}, both \\
radius buffer & ADD (\textbf{5\AA}, 10\AA), MUL ($\times$1.5, $\times$2.0) \\
MLP hidden scale & \textbf{1}, 2, 4 \\
dropout & 0.0, \textbf{0.1} \\
using layernorm & \textbf{Yes}, no \\
non linearities & \textbf{ReLU} \\
\midrule
\rowcolor[RGB]{234, 238, 234} \multicolumn{2}{l}{\textit{Training Config}} \\
learning rates & 1e-4, 7e-5, \textbf{5e-5}, 3e-5 \\
batch size & 8, \textbf{16} \\
pocket loss weight (cls-reg-radius) & \{0.5, \textbf{1.0}\}-\{\textbf{0.05}, 0.2\}-\{0.01, \textbf{0.05}, 0.2, 0.4\} \\
docking loss weight (coord-distmap-distill) & \{1.0, \textbf{1.5}, 2.0\}-\{0.0, \textbf{1.0}, 2.0\}-\{0.0, \textbf{1.0}\} \\
noise for predicted pockets & \textbf{range(0, 5)} \\
\bottomrule
\end{tabular}
\end{sc}
\end{small}
\end{center}
\end{table*}
\textbf{Training Details for \method{}.}
Our \method{} models are trained on eight NVIDIA Tesla V100 GPUs for 1500 epochs. We use Adam~\cite{kingma2014adam} as the optimizer, and set the hyperparameter $\epsilon$ to $1e-8$ and $(\beta_1, \beta_2)$ to $(0.9, 0.999)$ with no weight decay. The peak learning rate is set to $5e-5$ with a 15-epoch warmup stage followed by a linear decay learning rate scheduler. The dropout probability and total batch size are set to $0.1$ and $16$.

We list the detailed hyperparameter search configuration in Table~\ref{tab:fabind_hyperparameters}. For dynamic radius prediction, we try to aggregate embeddings from different components (protein embedding, ligand embedding, or both) and find that aggregating ligand embedding yields superior performance. Additive (ADD) and multiplicative (MUL) radius buffers are also tested. They have comparable performance, but the latter induces greater training instability; thus, we choose the additive buffer. The loss weights for each training objective prove to be sensitive in the multitask learning framework. Additionally, introducing noise to the three dimensions of the predicted binding site center during the training phase enhances model robustness significantly. Noises within a range of 0.0 to 5.0 Å are sampled each time.

\begin{table*}[htpb]
\caption[SearchSpace]{The hyperparameter options we searched through for confidence model. The final parameters are marked in \textbf{bold}.}
\label{tab:confidence_hyperparameters}
\begin{center}
\begin{small}
\begin{sc}
\begin{tabular}{lc}
\toprule
Parameter & Search Space  \\    
\midrule
\rowcolor[RGB]{234, 238, 234} \multicolumn{2}{l}{\textit{Model Config}} \\
model backbone & MLP, \textbf{stacked MLP}, node MLP, node Attention, FABind layer \\
MLP hidden scale & \textbf{1}, 4 \\
dropout & 0.1, \textbf{0.2} \\
\midrule
\rowcolor[RGB]{234, 238, 234} \multicolumn{2}{l}{\textit{Training Config}} \\
training objective & classification, ranking, \textbf{both} \\
learning rates & 1e-3, \textbf{1e-4}, 1e-5 \\
num of samples per batch & 8, \textbf{16} \\
num of copies & 4, \textbf{5}, 8 \\
\bottomrule
\end{tabular}
\end{sc}
\end{small}
\end{center}
\end{table*}


\textbf{Training Details for Confidence Model.}
The computational overhead for confidence model training is notably lower, requiring 15 epochs of training on eight NVIDIA Tesla V100 GPUs. The detailed hyperparameter options are listed in Table~\ref{tab:confidence_hyperparameters}. We evaluate various model backbones for feature extraction, namely MLP, Stacked MLP, Node MLP, Node Attention and FABind layer. Stacked MLP refers to the configuration where two previously described MLPs are stacked, thus expanding two sub-layers to four. Node MLP denotes an additional MLP transformation applied to node embeddings prior to their aggregation (sum). Subsequent to aggregation, another MLP updates the features for the output. Node Attention is similar to Node MLP but includes a self-attention module before aggregation. FABind layer results in the highest parameter count. However, we observe limited benefits in enhancing the confidence model's capability. This underscores that generalization is crucial for the efficacy of the confidence model.

During training, our confidence model can generate training examples on-the-fly, in contrast to the approach in DiffDock~\cite{corso2023diffdock}, which requires offline storage of sampled structures. This difference arises because DiffDock treats the scoring and confidence models as separate models, whereas our method integrates them into a single model. With the sampling mode on, the confidence model can directly utilize samples generated by \method{}. Consequently, an MLP-based backbone suffices due to the robust feature extraction capabilities of the preceding \method{}.

\section{More Visualizations}

\begin{figure}[htpb]
\centering
\includegraphics[width=0.99\linewidth]{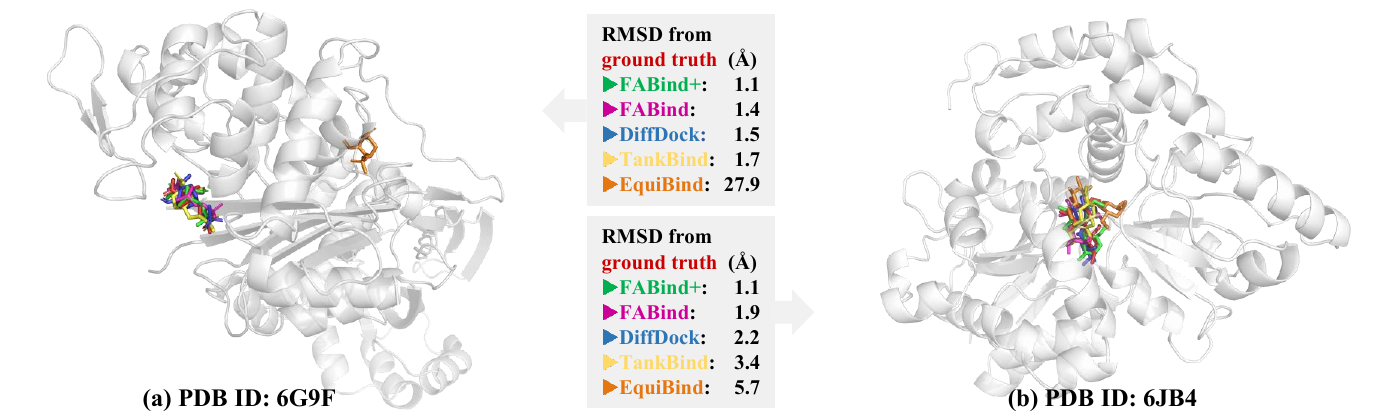}
\caption{Case studies for Regression-based \method{}. Structures predicted by FABind+ (green), FABind (magentas), DiffDock (blue), TankBind (yellow) and EquiBind (orange) are placed together with the protein target, with the RMSD to the ground truth (red) reported. These comparisons underscore the capability of \method{} for accurate prediction.}
\label{fig:ablation}
\end{figure}

\end{document}